\documentclass[amsmath,amssymb,12pt,floatfix] {revtex4}
\usepackage[dvips]{graphicx}
\usepackage{bm}
\begin{document}
\title{Diffusion of a hydrocarbon mixture in a  one-dimensional zeolite channel: an
exclusion model approach }
\author{Sakuntala Chatterjee and Gunter M. Sch\"{u}tz}
\affiliation{Institut f\"{u}r Festk\"{o}rperforschung,
Forschungzentrum J\"{u}lich, D-52425 J\"{u}lich, Germany}
\begin{abstract}

{\it Abstract:} Zeolite channels can be used as effective hydrocarbon traps. Earlier
experiments (Czaplewski {\sl et al.}, 2002) show that the presence of large
aromatic molecules (toluene)
block the diffusion of light hydrocarbon molecules
(propane) inside the narrow pore of a zeolite sample. As a result, the
desorption temperature of propane is significantly higher in the binary
mixture than in the single component case. In order to obtain further insight
into these results, we use a simple lattice gas
 model of diffusion of hard-core particles
 to describe the diffusive transport of two species of molecules in a 
one-dimensional
 zeolite channel. Our dynamical Monte Carlo simulations show that taking
 into account an
Arrhenius dependence of the single molecule diffusion coefficient
on temperature, one can explain many significant features
of the temperature programmed desorption profile observed in experiments.
However, on a closer comparison of the experimental curve and our simulation
data, we find that it is not possible to reproduce the higher propane current than toluene
current near the desorption peak seen in experiment. We argue that this
is caused by a violation of strict single-file behavior.
\end{abstract}
\maketitle

\section{Introduction}

Zeolites have wide industrial applications such as catalysts and adsorbents in
many chemical and petro-chemical processes.  
It has been argued that a
potential use for zeolites where adsorption and diffusion of
molecules become important is as hydrocarbon traps for the automobile
exhaust (see ~\cite{snurr} and the references therein).
When an automobile engine is turned on, the catalyst needs
a certain time to reach its
`light-off' temperature, which is around $250-300 ^{\circ} C $. This 
waiting time is
known as cold-start period. During this period, a considerable amount of
unburned hydrocarbon escape to the atmosphere through the tail-pipe. This loss
of fuel is
known as cold start problem. One possible solution to this problem is to use
some trap to adsorb the hydrocarbon molecules and release them once the
catalyst light-off has occurred. Zeolites are often used as effective
hydrocarbon traps in this regard. However, it is often found that while the larger and
heavier hydrocarbon molecules ({\sl e.g.} aromatics) are trapped successfully
inside the zeolite channels, the relatively light hydrocarbon components
 still manage to
escape before the combustion temperature has been reached. 

\begin{figure}
\includegraphics[scale=0.7,angle=0]{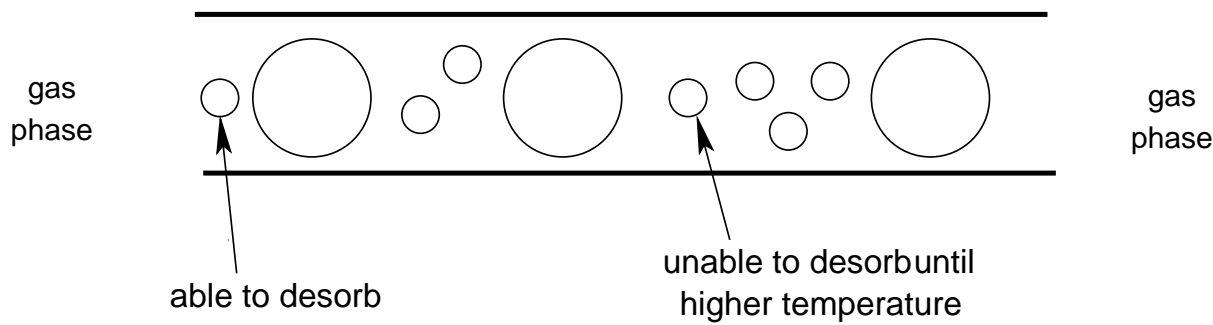}
\caption{\it Schematic diagram of $1$-d zeolite pore to show the trapping of
light, less-strongly adsorbed molecules by heavier, more-strongly adsorbed
ones. }
\label{fig:pore}
\end{figure}

In~\cite{snurr} Czaplewski {\sl et al.} have demonstrated that it is possible
to administer the controlled release of light hydrocarbon molecules in the
presence of large aromatic molecules using a one-dimensional zeolite channel.
They have argued that in certain zeolite samples, the channels are so narrow
that the hydrocarbon molecules cannot pass each other and the molecular
transport takes place in essentially a single-file condition.  In such
effectively one-dimensional zeolite channels, the
more strongly adsorbed heavy aromatic molecules block the motion of the less
strongly adsorbed light hydrocarbon molecules. 
 Because of the confining pore dimension, the light hydrocarbon
molecules cannot diffuse past the aromatic molecules (until higher
temperature) and are thus trapped inside the zeolite pore, as shown in 
Fig \ref{fig:pore}. As a result, the light
hydrocarbon molecules can desorb only
after the aromatic molecules have desorbed which occurs at a higher
temperature. This means that in the
presence of large aromatic molecules the light hydrocarbon molecules desorb at
a temperature which is much higher than the desorption temperature when no
aromatic molecules are present. 

 Czaplewski {\sl et al.} have performed an experiment using several zeolite
samples with varying pore dimensionality and studied the 
desorption profile of propane (light hydrocarbon) and toluene (heavy
aromatic molecule) mixture as the temperature is varied~\cite{snurr}. In
conformity with the argument presented in the previous paragraph, they have found
that inside a one-dimensional zeolite channel, the desorption temperature of
propane in a binary mixture is significantly higher than the single-component
desorption temperature. However, for a zeolite with three dimensional
connectivity, no such effect has been observed. This may be regarded
as experimental verification that single-file diffusion is
responsible for hydrocarbon trapping as described above.

It is the aim of this paper to provide a deeper theoretical understanding of the
experimental observation of single-file diffusion, as reported
in~\cite{snurr}. In particular, we are interested whether the assumption
of single-file behavior alone is sufficient to explain the experimental data. 
Several different approaches have
been used earlier to model the transport inside zeolites which include molecular
dynamics simulations, dynamical Monte Carlo simulation, transition-state theory
and Maxwell-Stefan approach [see~\cite{keil} for a detailed review on this
subject]. For the time-scales and length-scales that we are interested in,
the method of dynamical Monte Carlo simulation is the most suitable approach. 
To this end we follow the strategy of modeling diffusion in zeolites as introduced 
by K\"arger {\sl et al.}\cite{jk1,jk2}. 
In the context of investigating so-called molecular traffic control~\cite{mtc}
 they used dynamical Monte Carlo simulation of a stochastic lattice gas 
model for zeolite systems with a network of perpendicular sets of intersecting channels
(see \cite{BrSnano,BrS,RKS} for recent progress along these lines).
The basic idea of this model is to keep it as simple as possible with a minimum
of adjustable parameters. In this paper, we use a similar lattice gas model, 
first introduced in~\cite{chou,brzank2}, to describe  the
single-file diffusion of a two-component mixture of molecules in a one-dimensional 
zeolite channel.

To give some background, we note that a
 binary mixture adsorbed inside a zeolite channel has been
studied earlier using molecular dynamics simulation where the sorbate
molecules are assumed to interact with one another and with the zeolite
channel via a Lennard-Jones potential~\cite{snurr2}. The self-diffusivities  
of the two species were compared against experimental values obtained from
pulse field gradient NMR measurements. Both the simulations and experiments
show that as the fraction of the strongly adsorbed species was increased (at
constant total loading) the self-diffusivites of both components decrease 
(see also~\cite{santen}, where this concentration-dependence of single-file 
diffusivity is shown to affect the reaction kinetics inside a zeolite channel).
In the simulation in~\cite{snurr2},
 it was also found that the two different species show preferential
adsorption at two different portions of the zeolite channel which tentatively
supports the concept of molecular traffic control mentioned in the previous
paragraph (see also~\cite{clark}).

The attractive interaction between adsorbed molecules inside a zeolite pore is
found to give rise to important collective effects
 even when the molecules of only a single species
is present. In~\cite{sholl} the interaction between the neighboring adsorbate molecules was
modeled by a Lennard-Jones potential and the effect of the sorbate-pore
interaction was described by a sinusoidal potential. It was found that over a
wide temperature range, the fragmentation of molecular
clusters correspond to a higher energy barrier than the diffusion of a single
molecule, indicating that molecular clusters are stable. More importantly, the
energy barrier for cluster diffusion are often lower than that of a
single-molecule diffusion. These results show that the single-file diffusion
of molecules in a zeolite pore can be dominated by concerted movements of
molecular clusters.

We have not attempted any
detailed modeling of the interaction between sorbate molecules or between
sorbate and zeolite. Instead we have worked with a simple lattice gas model
and our Monte Carlo simulations show that within this simple model it
is  possible to explain major qualitative features of the experimental data of Czaplewski
 {\sl et al.}~\cite{snurr}. From this we conclude that only details of the experimental
curves are sensitive to details of the interactions among molecules and between 
molecules and the pore walls. However, it also transpires that the assumption
of strict single-file diffusion fails to reproduce one important property of 
the experimental
curves. This leads us to the further conclusion that some violation of the
single-file condition occurs.

 In the next section, we present a brief description of the
experimental setting of Ref. \cite{snurr} for the temperature programmed desorption (TPD)
profile for propane-toluene binary mixture in one-dimensional zeolite samples
with narrow (effectively, one-dimensional) pores. In section 3, we describe
our model in detail, followed by our results and discussions in the section
after.

\section{$\bm 1$-d zeolites as hydrocarbon traps: Experiment of Czaplewski {\sl
et al.}}

Czaplewski {\sl et al.} have considered several different zeolite samples with
different pore sizes and various pore network connectivity. To study zeolites with 
one-dimensional channels, EUO and Na-Mor were used and these data were compared 
against the data obtained for Na-ZSM-5, which is a zeolite with three-dimensional 
connectivity. In this section, we briefly summarise the features of
this series of experiments relevant for our study.

The zeolite samples were  loaded with an equimolar binary mixture of propane and
toluene and, for reference purposes, 
also with  single-component propane and toluene separately. After the
loading was complete, the whole system was purged in pure helium to remove all
hydrocarbon molecules from the gas-phase portion of the flow system (see Fig
\ref{fig:pore}). Then the
sample was heated at a constant rate and the outflow was monitored using a flame
ionisation detector and mass spectrometer. 

For a single-component loading,
it was found that as the temperature is steadily increased, the
instantaneous output current rises, attains a peak and then falls off. The
desorption temperature of each component was measured at the position of its
desorption peak. For the one-dimensional zeolite Na-MOR, single-component toluene
desorbs in two
different stages (showing two different peaks)
 and this induces a two-stage desorption for the propane in
the binary mixture. However, for the zeolite EUO, which also has a one-dimensional channel,
desorption takes place in a single stage. For this simplicity, we focus on EUO
and seek to describe its desorption profiles qualitatively using our model.

\begin{figure}
\includegraphics[scale=1.0,angle=0]{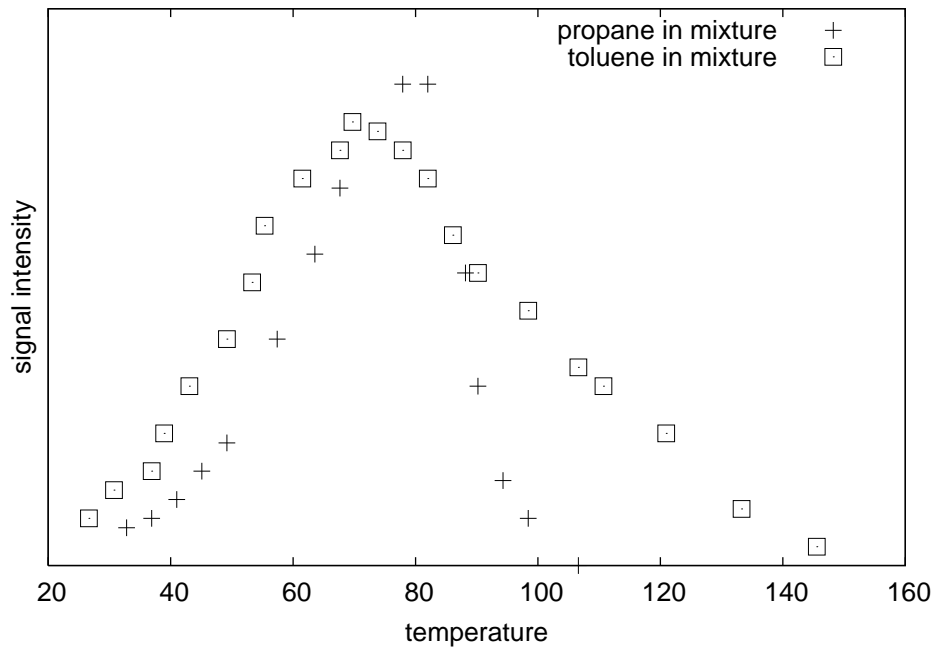}
\caption{Experimental data for TPD profile of propane and Toluene in binary
mixture in zeolite EUO, as measured by Czaplewski {\sl et al.} Data points
taken from Fig $5$ of~\cite{snurr}.  }
\label{fig:fig5}
\end{figure}

For the zeolite EUO, the single-component
propane desorption peak is found at $40 ^{\circ}$C and for single-component
toluene the peak occurs at  $80 ^{\circ}$C, toluene being more strongly
adsorbed. For an equimolar binary mixture of the two gases in EUO,
 the propane desorption peak is found to 
 occur at a substantially higher temperature ($75 ^{\circ}$C) and the toluene
desorbs at  $70 ^{\circ}$C, as shown in Fig \ref{fig:fig5}. 
 Note that toluene desorption precedes the propane
desorption, which suggests that in the narrow $1$-d channel of EUO the propane
molecules cannot pass the large toluene molecules and are able to desorb 
only after the toluene desorption peak has been reached.

This experiment demonstrates that using zeolite samples like EUO with $1$-d
channels, it is possible to trap light hydrocarbon molecules in presence of
large aromatic molecules, until high temperature. No such effect has been
observed for the $3$-d zeolite Na-ZSM-5, which is consistent with the absence 
of single-file diffusion in this material. In the next section, we
describe the lattice gas model that we use to explain the
experimental data for EUO qualitatively.

\section{Description of the Model}

\begin{figure}
\includegraphics[scale=1.0,angle=0]{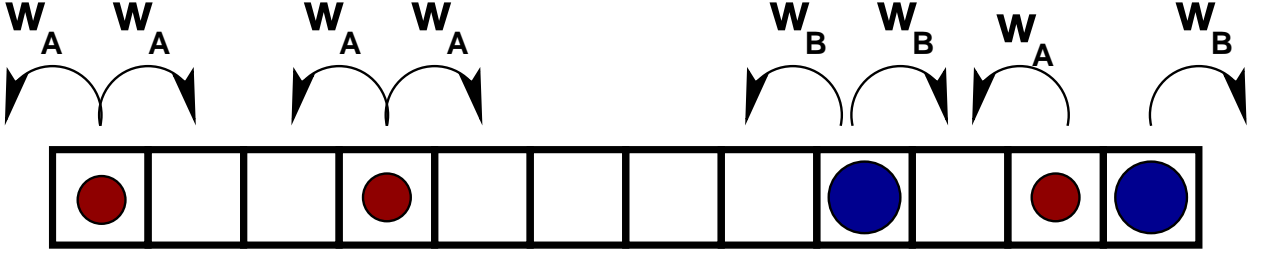}
\caption{\it Two component SEP on an open lattice}
\label{fig:lattice}
\end{figure}

In the spirit of the approach by K\"arger {\sl et al.}~\cite{jk1,jk2},
we model the narrow pores of EUO by a one-dimensional lattice whose
ends are open. In order to account for molecules of two different species 
with different diffusivities,
we extend their approach and introduce two kinds of particles as follows. 
The diffusion of propane and toluene in the pore is modeled
by a two-component symmetric exclusion process (SEP) on the lattice. We denote
a propane molecule by the symbol `$A$' and a toluene molecule by the symbol
`$B$'.
 A site can either be occupied by an $A$ particle or a $B$ particle or can
remain vacant (denoted by `$0$'). An $A$ or $B$ particle can jump with rate
$w_A$ or $w_B$, respectively, to the nearest
neighbor site on either side, if the site is empty. If the neighboring site is
occupied, no hopping is possible. The allowed moves are depicted in Fig \ref{fig:lattice}. 
Note that in~\cite{jk1,jk2}, it was assumed that $w_A = w_B$ but we use
different jump-rates (see Eq. \ref{eq:arr}) and this is one crucial aspect of
our model. 
The dynamical moves in the bulk are:

\begin{eqnarray}
\nonumber
A0 \stackrel {w_A}{\longrightarrow} 0A \\
\nonumber
0A \stackrel {w_A}{\longrightarrow} A0 \\
\label{eq:rate}
B0 \stackrel {w_B}{\longrightarrow} 0B \\
\nonumber
0B \stackrel {w_B}{\longrightarrow} B0   
\end{eqnarray} 

The $A$ or $B$ particles can exit through the boundary sites with rates $w_A$
and $w_B$, respectively. Therefore, the dynamical moves at the boundary are: 
\begin{eqnarray}
\nonumber
A \stackrel {w_A}{\longrightarrow} 0 \\
\label{eq:ratbdry}
B \stackrel {w_B}{\longrightarrow} 0 
\end{eqnarray}
Once they exit, they are removed from the system and
there is no boundary injection. Such a choice of boundary condition is 
motivated by the original experiment, where after the initial loading of the 
zeolite samples with hydrocarbon, in the course of the TPD measurement, 
hydrocarbons can desorb through the sample boundary, but no more hydrocarbons
are loaded into the sample.  

Eq. \ref{eq:rate} defines a two-component SEP which describes the diffusion of
two different species of hard-core particles. Note that when only a single particle 
$A(B)$ is
present in an otherwise empty lattice, then the system shows normal diffusive
behavior and the mean squared displacement (measured in units of
lattice-spacing) is proportional to the elapsed time (measures in units of Monte Carlo
step) with a proportionality constant $2w_{A\left ( B \right )}$. 
The two-component SEP was earlier studied
in~\cite{brzank2,brzank} for non-zero rates for boundary injection and
extraction, and constant values of $w_A$ and $w_B$. It was found
that the equilibrium distribution of particles is uncorrelated, {\sl i.e.} 
the probability that the $i$-th site contains an $A$($B$) particle
$Prob[n_i^{A(B)}=1]=\rho^{A(B)}$, independent of the occupancies of the other
lattice sites. Here $\rho^{A(B)}$ refers to the reservoir density and the
occupancy variable $n_i^{A(B)}$ is unity if $i$-th site is occupied by
$A$($B$), and $0$ otherwise. 

In this paper, we consider time-dependent hopping rates $w_A$ and $w_B$ to
model the TPD measurement carried out in experiment~\cite{snurr}.
As time goes on, the temperature $T$ of the
system is changed at a constant rate. 
The jump rates $w_A$ and $w_B$ are assumed to have the following 
Arrhenius dependence on
temperature~\cite{schuring}:
\begin{eqnarray}
\nonumber
w_A = \Gamma _ A \exp \left ( -E_A/kT \right ) \\
\label{eq:arr}
w_B = \Gamma _ B \exp \left ( -E_B/kT \right ) 
\end{eqnarray}

Here $k$ is the Boltzmann constant. Since Toluene is heavier, its
diffusivity should be less than propane, {\sl i.e.} $\Gamma_A > \Gamma_B$ and
$E_A < E_B$.   
Note that in this two-component SEP, hopping rates are explicitly
time-dependent which makes this problem difficult to deal analytically. In the
case of constant (time-independent) rates, one can use a set of coupled non-linear
differential equations to describe the time-evolution of the density profiles
and can study stationary state properties and the relaxation towards steady
state~\cite{brzank2,brzank}. One might expect that this 
Maxwell-Stefan type approach can be
extended to the time-dependent case as well, for very slow heating. However,
this requires the presence of local equilibrium---an assumption we find
difficult to justify in the present setting. Hence we use
dynamical Monte Carlo simulation to study the system. 


\section{Simulation Results on Two Component SEP }
\subsection{Dynamical Monte Carlo Simulation}
In this section, we present our results of dynamical 
Monte Carlo simulations on the above
lattice model. Assuming that the loading procedure generates an equilibrium
state (homogeneous bulk density),
we start from a random initial configuration drawn from the
equilibrium ensemble, where each site
can be occupied by an $A$ particles with probability $\rho_A$, by a $B$
particle with probability $\rho_B$, or remain vacant with probability
$(1-\rho_A - \rho_B)$. The system is then evolved following the dynamical
rules shown in Eqs. (\ref{eq:rate}), (\ref{eq:ratbdry}).
 The `temperature' $T$ of the system is
changed with time $t$ at a uniform rate:
\begin{equation}
T=T_0+\lambda t
\label{eq:temp}
\end{equation}
where $T_0$ is the initial temperature and $\lambda $ is the increment in temperature per
unit time. For each value of $T$, the jump rates are calculated from Eq. \ref{eq:arr}.

One Monte Carlo time-step consists of $(L+1)$ update
trials. At each update trial, we choose a lattice bond at random. If
the bond lies in the bulk, the occupancies of the pair of sites adjacent to
the bond are updated according to the rules given in Eq. \ref{eq:rate}: if one
of these two sites contains an $A$($B$) and the other is empty, then a hopping
across the bond takes place with probability $w_A$($w_B$). To perform this
process, we draw a
random number from a uniform distribution in the range $[0,1]$ and if the
number is less than $w_A$($w_B$), then we exchange the occupancies of the
pair of sites. If both the adjacent sites of the selected lattice bond are
occupied, no action is taken. 
If the chosen lattice bond connects the boundary site 
(in this case, the leftmost or the rightmost site of the 
lattice) to the reservoir, then desorption takes place:  
if the boundary site is occupied by an
 $A$($B$) particle, then the site is emptied with a probability $w_A$($w_B$),
following the same steps discussed above.    

 We measure the number of particles of each species coming out  
through the boundary sites per unit time, as a function of temperature (or
time). We
call this quantity instantaneous current and denote as $J_A(t)$ and $J_B(t)$,
for outflow of $A$ and $B$, respectively.

\subsection{Single component loading}
When a single species, either $A$ or $B$ is present, then the system executes
a single-component SEP with time dependent rates. As temperature increases,
the diffusivity increases and more and more particles come out from the
lattice through the boundary sites. As a result the current increases. But
since there is no boundary injection, the system starts getting depleted of
particles  and the current finally decreases after reaching a
peak, as found in experiment~\cite{snurr}.

Before we present our simulation results, a note about the
 choice of parameters is
in order. Our aim is to provide a qualitative understanding of the
experimental data. Within our simplistic model we do not expect an exact
quantitative agreement with experiment. We have carried out simulation within
the same temperature range as considered 
in the experiment and have chosen the activation
energies and temperature increment rate $\lambda$ such as to obtain a desorption peak
within this temperature range. However, the position of the desorption peak in
our simulation does not match with the experimentally observed value. 

In the experiment diffusion of propane and toluene was studied inside an EUO
pore. Typical channel length of an EUO zeolite is $5$ $\mu m$ and the
molecular diameters of propane and toluene are, respectively, $\sim 4.4$
$\AA$ and  $\sim 5.7$ $ \AA$. The ratio of the channel length to the
molecular size is $\sim 1000$, which should correspond to the number of
lattice sites
in our model. However, since our focus is to explain the main
experimental results qualitatively, we work with a smaller lattice for
computational efficiency. Also, the amount of substance adsorbed 
inside an EUO pore depends on the details of the structure of
the pore and the adsorbed species. In our model, we have not taken into
account of such details and have worked with an intermediate value of the
density of $A$ and $B$ particles in the lattice. However, we have observed
that even when a different value of the density is chosen, there is no qualitative
change in our data (details not presented).

\begin{figure}
\includegraphics[angle=0,scale=1.0]{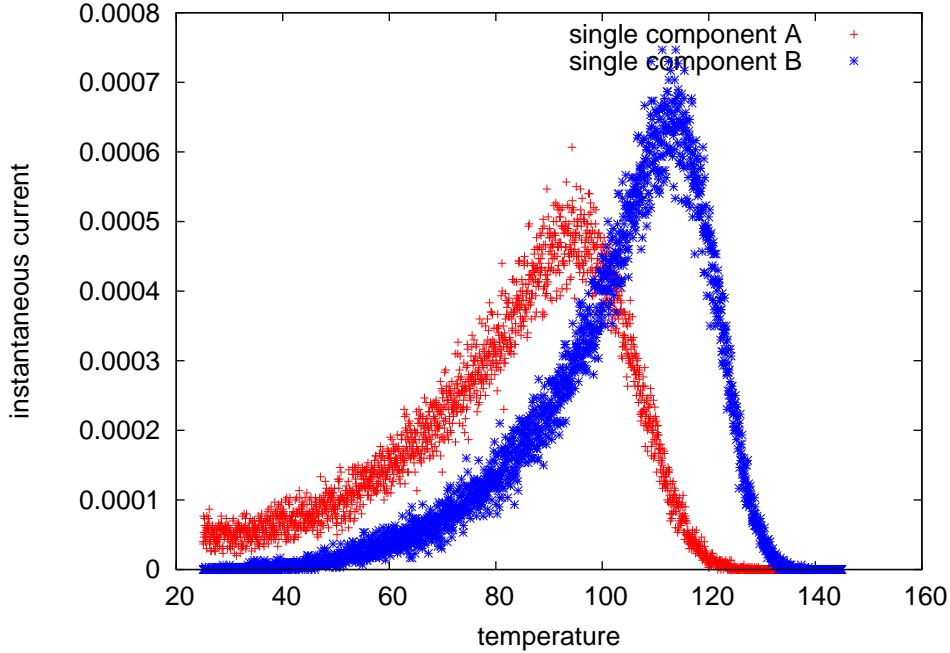}
\caption{\it Variation of instantaneous current as a function of temperature
 (in $^{\circ}$C) for single-component loading.}
\label{fig:single}
\end{figure}

In our simulation, we have used a lattice of length $L=100$. In the initial
configuration, each site contains
an $A$ particle with a probability  $\rho_A=0.4$. At the end of each Monte
Carlo time-step, temperature is incremented by $5 \times 10^{-4}$ degree,
 starting
from an initial value of $27 ^\circ C$ to a final value $150 ^ \circ C$.  
 We have used
$E_A=83.1$ kJ/mol. In order to make our simulation more efficient, the
factor  $\Gamma_A$ which sets the Monte Carlo time-scale,
has been given a large value such that
the variation of $w_A$ 
in the above temperature range is substantial. This is
ensured by setting $\Gamma_A = \exp \left (E_A/kT_f \right ) $, where $T_f$ is
the final temperature. To obtain good statistics, we have averaged over
$300,000$ initial conditions.  We present our data for $J_A(t)$ in Fig
\ref{fig:single}. In the same figure, the data for single-component loading of
$B$ molecules is also shown, with $E_B=124.7$ kJ/mol,
 $\rho_B=0.4$ and $\Gamma_B=\exp \left (E_B/kT_f \right )$.   
Note that the desorption peak for $J_B(t)$ occurs at a higher temperature than
that for  $J_A(t)$. In other words, $B$ is more strongly adsorbed than $A$, as
expected.

\subsection{Loading of the binary mixture}
\begin{figure}
\includegraphics[scale=1.0,angle=0]{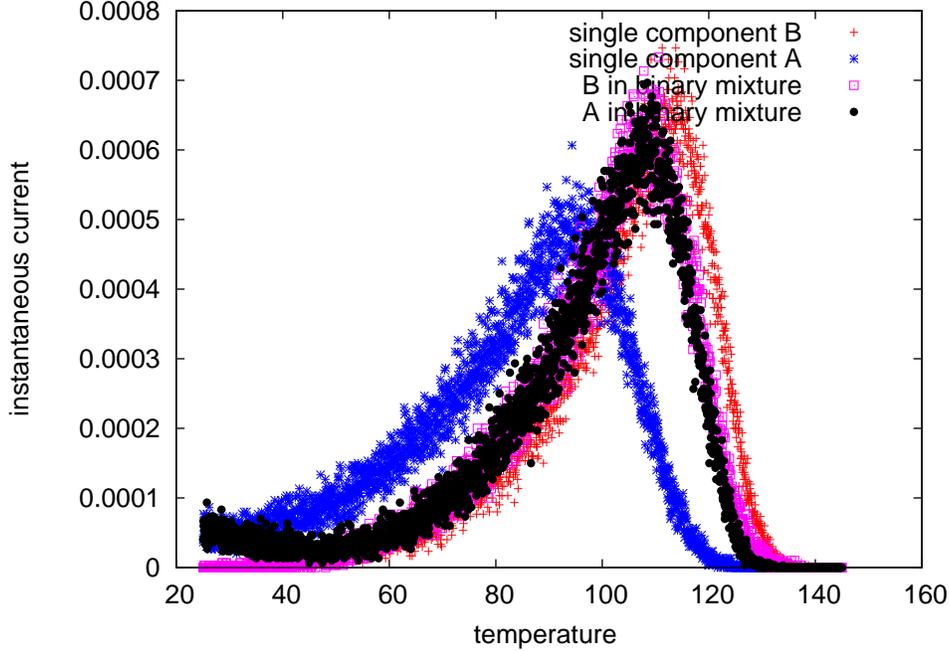}
\caption{\it Plot of current vs temperature (in $^\circ$C)
for a binary mixture of $A$ and
$B$. We have used $L=100$, $\rho_A=\rho_B=0.4$, $\lambda =5 \times 10^{-4}$  
$^{\circ}$C per unit time,
 $E_A=83.1$ kJ/mol, $E_B=124.7$ kJ/mol, $\Gamma_A=exp (E_A/kT_f)$ and 
$\Gamma_B=exp (E_B/kT_f)$. The data has been averaged over $300,000$ initial
configurations. For comparison, single-component data presented in 
Fig \ref{fig:single} has been replotted.}
\label{fig:twopeak}
\end{figure}

In the original experiment, zeolite samples were loaded with equimolar
mixtures of propane and toluene. Likewise, we start with an initial
configuration with $\rho_A = \rho_B$. 
Starting with a uniform mixture of $A$ and $B$, the
plot of $J_A(t)$ and  $J_B(t)$ is shown in Fig \ref{fig:twopeak}
 against the variation of temperature.   

We find that in the binary mixture, the positions of the peak of $J_A(t)$ and
$J_B(t)$ almost coincide. The desorption temperature of $B$ 
is close to that in the single-component case, but desorption
temperature of $A$ is substantially higher than the single-component case.  
This clearly shows 
that the presence of $B$ strongly influences the desorption of $A$. 
 However, as seen from
our data, the outflow of $A$ particles at low temperature is not suppressed
much by the presence of B. This means that a considerable fraction of $A$
particles still manages to escape at lower temperature.
 This happens due to the presence of $A$ molecules
in the boundary region of the 
lattice which are not trapped by any $B$ molecule.
 When we start with a uniform mixture of $A$ and $B$ as the
initial condition, there can be few $A$ molecules at the boundary, which
do not have any $B$ molecules blocking their exit through the boundary.

\begin{figure}
\includegraphics[scale=1.0,angle=0]{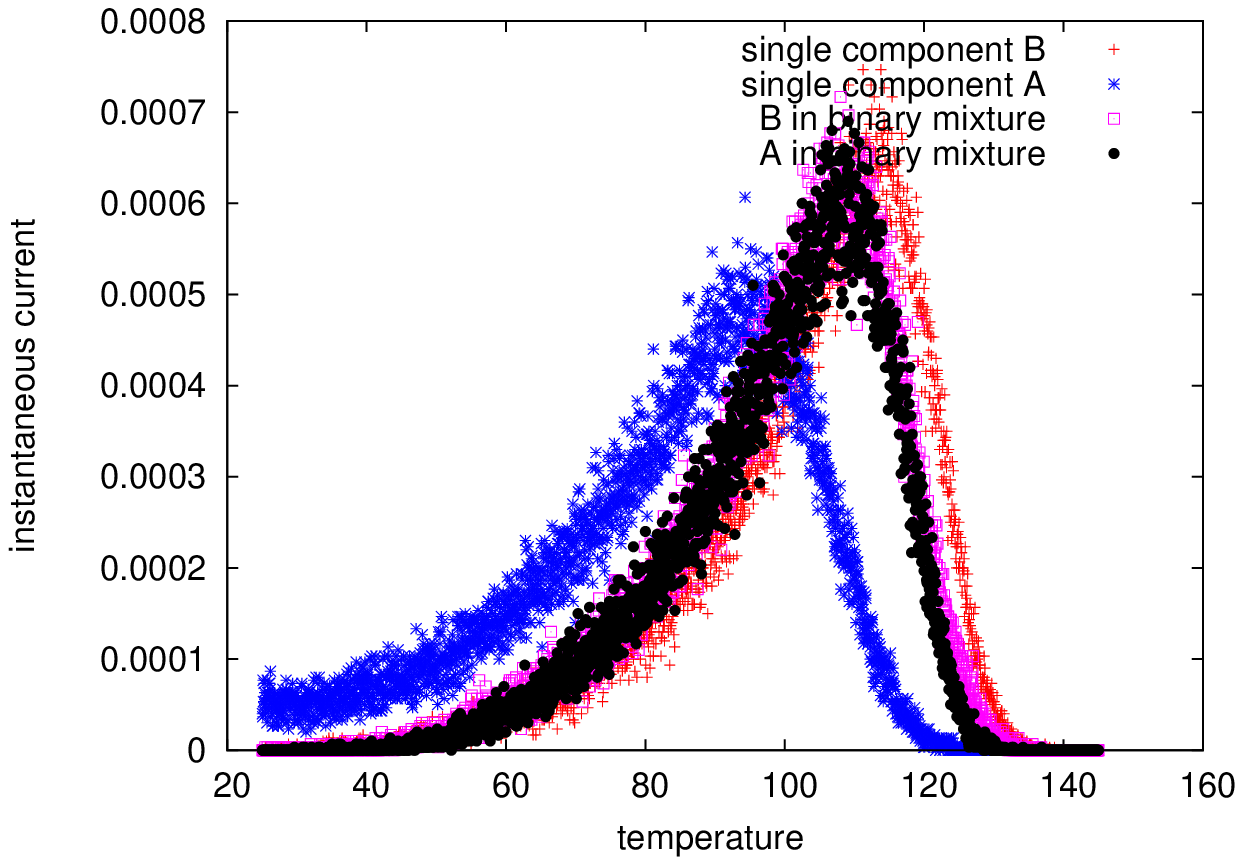}
\caption{\it The low temperature outflow of $A$ particles, as shown in
in Fig \ref{fig:twopeak}
is suppressed when all untrapped $A$
particles are removed from the initial configuration. The other simulation
parameters are same as in Fig \ref{fig:twopeak}.}
\label{fig:purge}
\end{figure}

We verify that if all such untrapped $A$ molecules are removed from the initial
configuration, then the low temperature outflow of $A$ is suppressed (see Fig
\ref{fig:purge}). 
We would like to comment that in the original experiment, after the zeolites
were loaded with propane and toluene mixture, the system was passed through a
`gas-purging' phase~\cite{snurr}. In this phase, all the molecules were
removed from the `gas phase' (shown in Fig \ref{fig:pore}). It is very likely
that the weakly adsorbed propane molecules which are untrapped by toluene
molecules  and reside
close to the boundary sites, are also purged out of the system at this stage.
So our choice of initial configuration with no untrapped $A$ particle, is
not unrealistic but resembles the actual experimental scenario.

\subsection{Relaxing the single-file condition}
If we closely
examine the experimental data shown
 in Fig \ref{fig:fig5}, then we find that near the
peak of the desorption profile the propane current is higher than the toluene
current. The propane desorption peak is about $8 \%$ higher than that of
toluene. However, in a strictly single-file condition, where propane molecules can
 escape only after the toluene molecules have desorbed,   propane current can
never exceed toluene current if we start from an equimolar mixture of the two
[Fig \ref{fig:purge}].

\begin{figure}
\includegraphics[scale=0.5,angle=0]{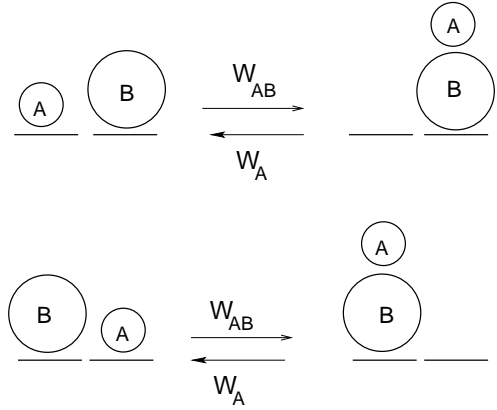}
\caption{\it $A$ particles can cross over $B$ particles with a small rate
when the single-file condition is relaxed.}
\label{fig:cross}
\end{figure}

 Therefore, to explain the experimental observation, 
we have to modify our model of the exclusion process. 
To produce a higher propane current, as seen in experiment,
 we propose to slightly relax the single-file condition.    
In our model, 
we allow an $A$ particle to diffuse past a $B$
particle with a small rate, as shown in Fig \ref{fig:cross} where
\begin{equation}
w_{AB} = \Gamma_{AB} \exp \left (- E_{AB}/kT \right ). 
\label{eq:cross}
\end{equation}
The other moves remain the same as in Eq. \ref{eq:rate}. Here, we choose 
$E_A < E_B <E_{AB}$ such that the violation of single-file condition becomes
important only at sufficiently high temperature.

\begin{figure}
\includegraphics[scale=1.0,angle=0]{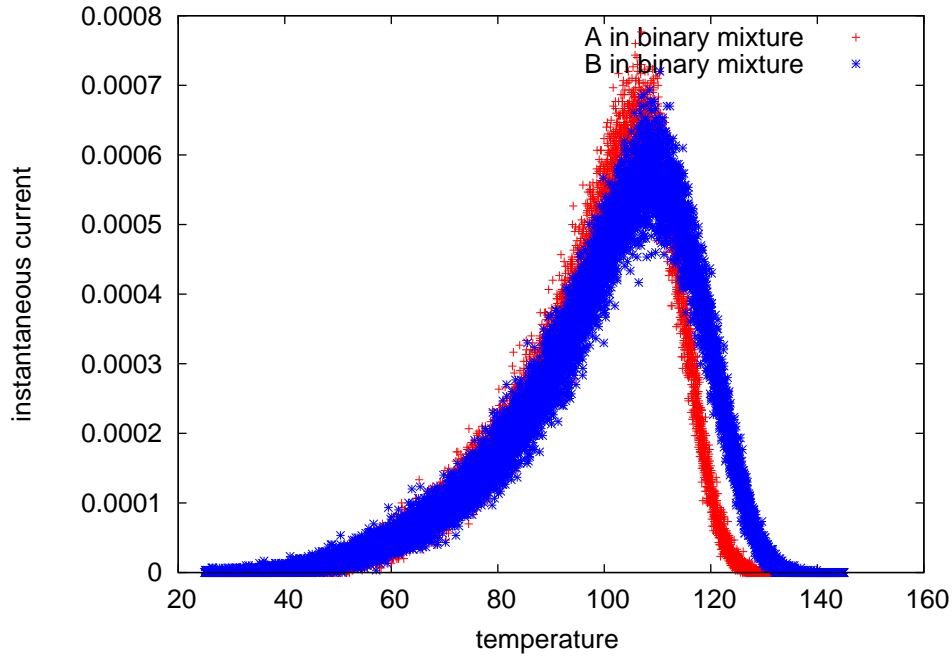}
\caption{\it Desorption profile for $A$ and $B$ with a relaxed single-file
condition. We have used $E_{AB}=166.2$ kJ/mol and 
$\Gamma_{AB}=exp(E_{AB}/kT_f)$. The other simulation parameters remain same as
 in Fig \ref{fig:twopeak}.The data has
been averaged over $300,000$ `gas-purged' initial conditions.}
\label{fig:wide}
\end{figure}

We present our simulation
results with this modified model in Fig \ref{fig:wide}. We obtain a larger $A$
current than $B$ current near the peak, as observed experimentally. Our data
shows that the desorption peak of $A$ is about $7 \%$ higher than that of $B$,
close to the corresponding experimental value of $8 \%$. 
 It is useful to have a 
rough estimate of the relative magnitudes of the different hopping
probabilities $w_A$, $w_B$ and $w_{AB}$ in our simulation.
 With our choice of parameters, near the
desorption peak, the ratio $w_{AB}/w_B$ attains a value of about $0.3$ and
$w_{AB}/w_A$ is close to $0.09$. We have found that the fraction of $A$ particles that
desorb after crossing over a $B$ particle at the boundary, constitutes about $2.5 \%$ 
of the total current of $A$ particles, at the desorption peak. Hence the $A$ current has
a rather small contribution from the crossing events at the boundary, but the 
desorption profile is nevertheless significantly affected by such crossings (see Fig 
\ref{fig:wide}).      

\section{Conclusion}
It is important to understand the transport
mechanism of the molecules within the pores of zeolite in order to design more
efficient applications for these materials. 
In this paper, we have used a two-component SEP to model the diffusion of
propane and toluene in an effectively one-dimensional zeolite channel. Within
this simple lattice gas model we have been able to explain the qualitative
features of the temperature dependent desorption profile observed in
experiment~\cite{snurr}. Our model for two-component single-file diffusion
shows that the presence of strongly adsorbed $B$ particles raises the effective
desorption temperature of weakly adsorbed $A$ particles, as seen in the
original experiment. 
This indicates that no detailed modeling of the interaction between the
molecules or between the molecule and the pore-wall and  boundary
effects are required to explain the main experimental results qualitatively.

Our Monte Carlo simulation shows that in order to
reproduce  a propane current larger than the toluene current, one has to relax
the single-file condition and allow the particles to cross each other 
for sufficiently high temperature.  A violation of the
single-file condition in zeolite channels has been found earlier while
studying tracer diffusion of methane molecules in $AlP0_4$-$5$
zeolite using molecular dynamics simulation~\cite{tepper}. The long-time growth of the
mean squared displacement was found to be linear, as in normal diffusion. It
was argued that because of narrow channels in $AlP0_4$-$5$, the methane
molecules are not able to cross each other easily, but infrequent crossings do
take place. This gives rise to a diffusive regime on large time-scales. On
short time-scales, an intermediate behavior was found where mean squared
displacement is proportional to $t^{0.64}$. These results are in contrast
with the experimental observation of single-file diffusion in
$AlP0_4$-$5$~\cite{kukla}. In order to gain a better insight into these
findings, a simplified ``hop and cross" model was introduced
that correctly reproduced the molecular dynamics results~\cite{tepper}.
In this ``hop and cross" model, a particle jumps to the adjacent site with a
probability $p_{hop}$ if the site is empty. If the neighboring site is occupied, then
the two particles interchange their positions with a probability $p_{cross}$. In our
model, where two different species are present, we have introduced a similar
crossing effect by allowing the smaller particle to climb over the larger one
(see Fig \ref{fig:cross}).  We have found that near the desorption peak,
the $A$ current has a rather small contribution from the
crossing-events (about $2.5 \%$). However, such crossings do play a crucial
role and  affect the desorption profile, as seen in Fig \ref{fig:wide}.
 It would be
interesting if it is possible to estimate the fraction of such crossing-events from
the relative heights of the two peaks. However, we do not have much insight
into this at present.

Throughout, we have assumed that the temperature dependence of the
diffusivities $w_A$, $w_B$ and $w_{AB}$ follows the Arrhenius law (see 
Eqs. (\ref{eq:rate}), (\ref{eq:cross})). This may not
hold true. In fact for certain zeolites, non-Arrhenius temperature dependence
has been discussed in earlier studies~\cite{schuring}.  However, to our knowledge, such
behavior for EUO zeolite has not been reported. Besides, our simple
Arrhenius dependence yields good qualitative agreement with the experiment.
Also, no qualitative difference is observed when the initial density
is changed. Thus, 
within our model, we do not find any strong dependence on density. This
conclusion is open to experimental verification.

In our model, we have not considered any interaction between the molecules,
except hard-core exclusion. Since we can explain the main features of the
experimental data qualitatively, it seems that the interaction 
details between the molecules do
not play a crucial role. However, earlier studies have shown that these
interactions can give rise to important collective effects~\cite{sholl} and
hence might be relevant for a
quantitative comparison with the experiment. 
In our model, we have also neglected any interaction
between the molecules and the pore-wall. In particular, for an EUO zeolite, it
is known that the narrow channel is accompanied by large
side-pockets~\cite{pocket}. These
side-pockets might give rise to a non-trivial interaction between the
pore-wall and the molecules which in turn might affect the distribution for
the residence-time of the molecules~\cite{res1,res2}. It would be of interest to see what
happens to the simulated desorption profile once the effect of the pore-wall
interaction has been
incorporated into our model.

\section{Acknowledgments}
Financial support by the Deutsche Forschungsgemeinschaft within the priority
 programme SPP1155 is gratefully acknowledged. We would like to thank J.
 K\"arger for inspiring discussions on single-file diffusion and diffusion in 
zeolites.

\end{document}